\documentclass{PoS}

\title{An Overview of Maximal Unitarity at Two Loops}

\ShortTitle{An Overview of Maximal Unitarity at Two Loops}

\author{Henrik Johansson\\
        Theory Group, Physics Department, CERN\\
         CH--1211~Geneva~23, Switzerland}

\author{\speaker{David A. Kosower}\\
       Institut de Physique Th\'eorique, CEA--Saclay,\\
          F--91191 Gif-sur-Yvette cedex, France}

\author{Kasper J.~Larsen\\
            NIKHEF, Science Park 105,\\
            NL--1098~XG Amsterdam, The Netherlands}

\abstract{We discuss the extension of the maximal-unitarity method to 
two loops, focusing on the example of the planar double box.  Maximal cuts
are reinterpreted as contour integrals, with the choice of contour fixed by 
the requirement that integrals of total derivatives vanish on it.  
The resulting
formul\ae{}, like their one-loop counterparts, can be applied either 
analytically or numerically.}

\FullConference{Loops and Legs in Quantum Field Theory - 11th DESY Workshop on Elementary Particle Physics,\\
		April 15-20, 2012\\
		Wernigerode, Germany}

\def\Ord{{\cal O}}
\def\qbar{\bar q}
\def\NeqFour{{\cal N}=4}
\def\e{\epsilon}
\def\spa#1.#2{\left\langle#1\,#2\right\rangle}
\def\spb#1.#2{\left[#1\,#2\right]}
\def\spash#1.#2{\left\langle\smash{#1}\,\smash{#2}\vphantom{1}\right\rangle}
\def\spbsh#1.#2{\left[\smash{#1}\,\smash{#2}\vphantom{1}\right]}
\def\sand#1.#2.#3{%
\left\langle\smash{#1}{\vphantom1}^{-}\right|{#2}%
\left|\smash{#3}{\vphantom1}^{-}\right\rangle}
\def\sandpm#1.#2.#3{%
\left\langle\smash{#1}{\vphantom1}^{+}\right|{#2}%
\left|\smash{#3}{\vphantom1}^{-}\right\rangle}
\def\sandpp#1.#2.#3{%
\left\langle\smash{#1}{\vphantom1}^{+}\right|{#2}%
\left|\smash{#3}{\vphantom1}^{+}\right\rangle}
\def\sandb#1.#2.#3{%
\left\langle\smash{#1}{\vphantom1}^{\pm}\right|{#2}%
\left|\smash{#3}{\vphantom1}^{\pm}\right\rangle}
\def\spar#1..#2{\langle\!\langle#1\cdots#2\rangle\!\rangle}
\def\Global{{\cal G}}
\def\jig#1{J_{\oint_{}^{}\!,#1}}
\def\treelevel{{(0)}}

\begin{document}

\section{Introduction}

Amplitudes are the basic building blocks for physics predictions in
QCD.  Predictions of differential cross sections are essential to
controlling backgrounds to new physics at the Large Hadron Collider
(LHC).  Because of their strong dependence on the unphysical
renormalization and factorization scales, leading-order (LO)
predictions are not quantitatively reliable.  Next-to-leading order
(NLO) calculations give the first quantitative predictions of
processes involving QCD.  NLO calculations require one-loop
amplitudes, in addition to other ingredients.  Recent years have seen
a revolution in our ability to calculate these amplitudes, thanks to
maximal unitarity~\cite{Zqqgg,BCFUnitarity} and other
developments~\cite{Revolution} such as the
Ossola--Papadopoulos--Pittau~\cite{OPP} decomposition.

For some processes, such as $gg\rightarrow W^+ W^-$ and $gg\rightarrow
ZZ$, the tree-level amplitude vanishes, and accordingly the one-loop
amplitude furnishes only an LO prediction.  Such subprocesses are
nominally higher order in the strong coupling, of $\Ord(\alpha_s^2)$,
compared to $\Ord(\alpha_s^0)$ for the basic $q\qbar\rightarrow W^+
W^-, ZZ$ subprocesses. This is however partly compensated by the much
larger gluon densities, so that they merit computation.  To compute
these subprocesses to NLO, one needs two-loop amplitudes.

Two-loop amplitudes are also needed for NNLO calculations, which in
turn will be needed for future precision physics at the LHC.  Such
calculations will also be useful in providing honest uncertainty
estimates for existing NLO predictions.

\section{On-Shell Methods}

Traditional Feynman-diagrammatic methods suffer from an explosion in
the number of diagrams, and an even greater explosion in the number of
terms, as the number of external legs or the number of loops
increases.  Yet many results for amplitudes, especially in the
$\NeqFour$ supersymmetric gauge theory, are extremely compact; and all
known loop results in gauge theories are vastly more compact than
would be suggested by the number of diagrams.  This reflects the vast
redundancy present in Feynman diagrams, due to 
explicit handling of 
non-physical states, and the resulting gauge dependence of intermediate
quantities.  On-shell methods use only information from
physical, on-shell, states to compute amplitudes, thereby avoiding
throughout the computation of gauge-variant quantities which must
cancel at the end.  This makes calculations simpler and made possible
new NLO calculations at high multiplicity, such as those of
$W,Z\,+\,4$~jets and $W\,+\,5$ jets~\cite{VplusManyJets}.

On-shell methods make use of general properties of amplitudes to
derive tools for computations: tree-level factorization leads to the
Britto--Cachazo--Feng--Witten
on-shell recursion relations for tree 
amplitudes~\cite{BCFW}; the unitarity
of the $S$-matrix gives rise to the unitarity~\cite{BasicUnitarity}
and generalized-unitarity methods; and the presence of an underlying
field theory allows for a representation in terms of an integral
basis.  The formalism can be summarized in the following equation,
\begin{equation}
\textrm{Amplitude} = \sum_{j\in \textrm{Basis}} c_j \textrm{Int}_j + \textrm{Rational}\,,
\label{BasicEquation}
\end{equation}
where the sum is taken over a basis of integrals, and the coefficients
$c_j$ as well as the remaining rational terms are rational functions
of spinor variables.  For analytic calculations, having a basis of
linearly-independent integrals simplifies calculations but is not
strictly essential.  For numerical calculations, it is essential.

\section{Integral Basis}

\begin{figure}[t]
\centering
\includegraphics*[scale=0.6]{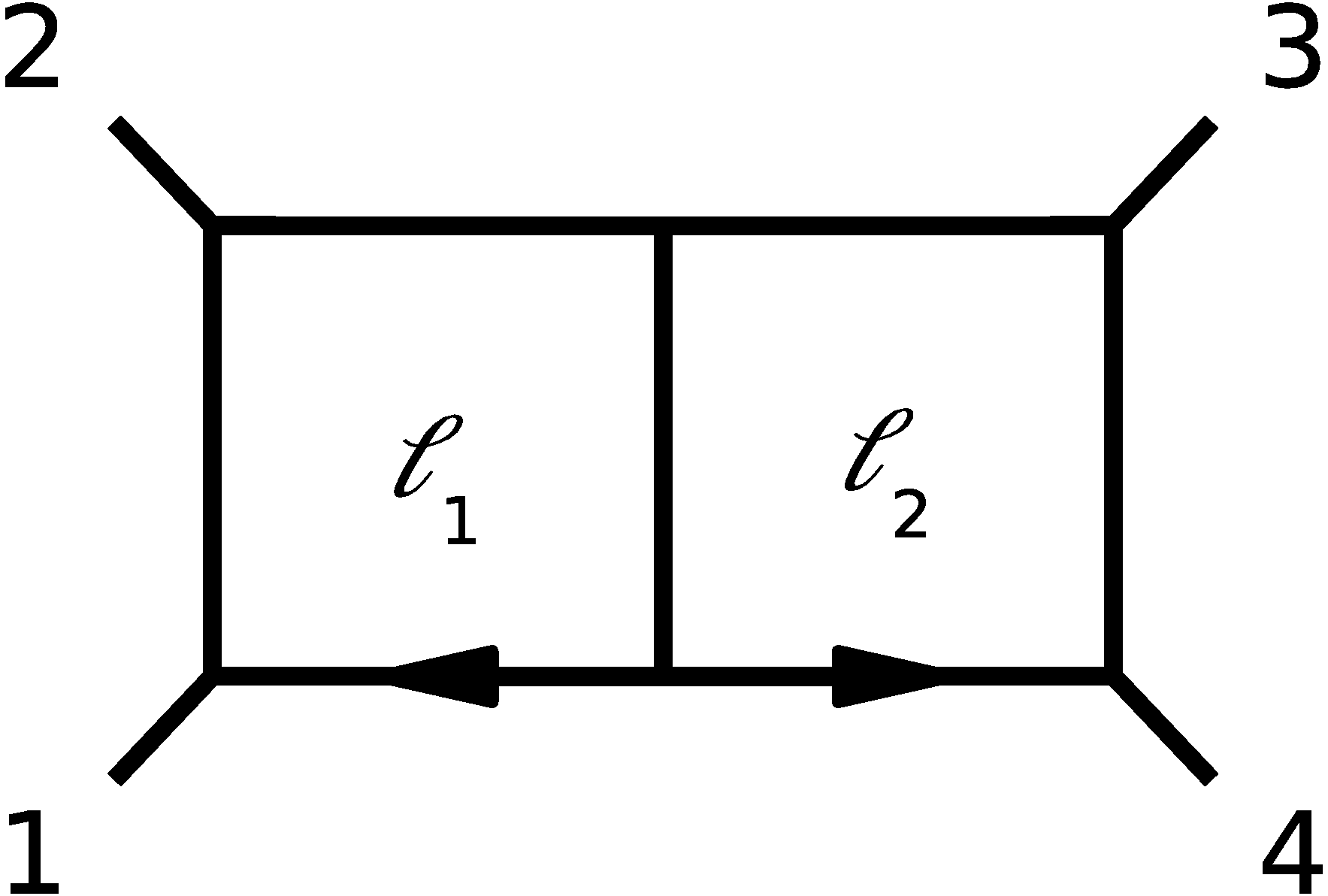}
\caption{The double-box integral
\label{DoubleBoxFigure}}
\end{figure}

At one loop, the basis for computations in massless gauge theories
consists of box, triangle, and bubble integrals, where all internal
lines are massless, and external legs may be massive or massless.  In
general, we must distinguish between two different notions of basis: a
``$D$-dimensional basis,'' which keeps terms to all orders in the
dimensional regulator $\e$, and a ``regulated four-dimensional
basis,'' which keeps only terms through $\Ord(\e^0)$.  Because some
integrals are linearly independent only at $\Ord(\e)$, the latter
basis is more compact.  We will implicitly be using this latter basis
in these Proceedings.  The planar part of the basis at two loops will
contain integrals with up to eight propagators~\cite{TwoLoopBasis}.
Beyond one loop, some basis integrals will necessarily contain
irreducible numerators, numerators which cannot be written as linear
combinations of inverse propagators.  For the double boxes we
consider here and shown in fig.~\ref{DoubleBoxFigure}, for example,
there are two basis integrals when all external legs are massless, or
when one external leg is massive.

\section{Maximal Unitarity}

In the basic unitarity method at one loop, we sew together two tree
amplitudes with a phase-space integral, and promote the
positive-energy on-shell delta functions to off-shell propagators.  The
resulting object will yield an integral containing the correct
contribution to the target amplitude in the channel in which we
have performed the sewing.  We must still reduce the resulting
integral symbolically in order to separate the contributions from
different basis integrals, and find their respective coefficients.
Finally, we must merge contributions from all channels.  This sewing
procedure inverts the procedure of cutting a one-loop amplitude, in
which we replace a pair of propagators surrounding the given channel
by positive-energy delta functions.  This isolates all integrals
containing those two propagators.  There are of course many such
integrals: various boxes and triangles, and a bubble integral as well.

In order to isolate a smaller number of integrals, we must cut more propagators.  This is possible; indeed at one loop,
we have four degrees of freedom in the loop momentum (ignoring the $(-2\e)$-dimensional components), and so we
can imagine cutting four propagators at once~\cite{BCFUnitarity}.  There is, however, a subtlety involved.  We might imagine replacing
the four propagators in the box integral,
\begin{equation}
\int {d^{4-2\e}\ell\over (2\pi)^{4-2\e}}\;
{1 \over \ell^2 (\ell-k_1)^2 (\ell-K_{12})^2 (\ell+k_4)^2}\,,
\end{equation}
by positive-energy delta functions,
\begin{equation}
\int {d^{4-2\e}\ell\over (2\pi)^{4-2\e}}\;
{\delta^{(+)}\bigl(\ell^2\bigr)\delta^{(+)}\bigl((\ell-k_1)^2\bigr)
 \delta^{(+)}\bigl((\ell-K_{12})^2\bigr)
\delta^{(+)}\bigl((\ell+k_4)^2\bigr)}\,.
\end{equation}
These delta functions instruct us to solve the simultaneous equations,
\begin{equation}
\ell^2 = 0\,,\quad
-2\ell\cdot k_1+k_1^2 = 0\,,\quad
-2\ell\cdot k_{2}+K_{12}^2-k_1^2 = 0\,,\quad
2\ell\cdot k_4+k_4^2 = 0\,,
\label{QuadCutEquations}
\end{equation}
which are linear combinations of the delta-function arguments.
Let us examine the special case when legs~1, 2, and~4 are massless; we can then solve the first, second,
and last equations by setting,
\begin{equation}
 \ell^\mu = \frac{\xi}2 \sand1.{\mu}.4\,,
\end{equation}
and then solve for $\xi$,
\begin{equation}
\xi = -\frac{\spb1.2}{\spb2.4}\,,
\end{equation}
using the third equation in eq.~(\ref{QuadCutEquations}).
Similarly, we see that there is a second solution,
\begin{equation}
\ell^\mu = -\frac{\spa1.2}{2\spa2.4}\sand4.{\mu}.1\,.
\end{equation}

The subtlety arises from the fact that for generic external momenta,
these solutions are complex.  The domain of the delta functions, on
the other hand, is real; taken literally, the delta functions would
yield zero!  Cutting both sides of eq.~(\ref{BasicEquation}) would
then give us the equation $0=0$, which is true but not very useful.
This is the same issue that arises in straightforward interpretations
of delta functions in the connected picture for twistor-string
amplitudes~\cite{RSV}.

To find a solution to this subtlety, we may note~\cite{Vergu} that
contour integration behaves very much like integration over a delta
function,
\begin{equation}
\oint_{C(z_0)} \hskip -2mm dz\; \frac{{\rm Poly}_1(z)}{{\rm Poly}_2(z)-a}
= \frac{{\rm Poly}_1(z_0)}{{\rm Poly}_2'(z_0)}\,,
\label{ContourIntegral}
\end{equation}
where $z_0$ is defined by the equation ${\rm Poly_2}(z_0)=a$ (with multiciplicity one).
  We could {\it define\/} the desired delta function
as follows,
\begin{equation}
\int dz\; {\rm Poly_1}(z)\delta({\rm Poly}_2(z)-a)
\equiv \oint_{C(z_0)} \hskip -2mm dz\;
\frac{{\rm Poly}_1(z)}{{\rm Poly}_2(z)-a}\,.
\end{equation}
There is one significant difference from ordinary delta function
integration: there is no absolute value around the derivative in the
denominator on the right-hand side
of eq.~(\ref{ContourIntegral}), so that the result remains an analytic function.

\begin{figure}[t]
\centering
\includegraphics*[scale=0.5]{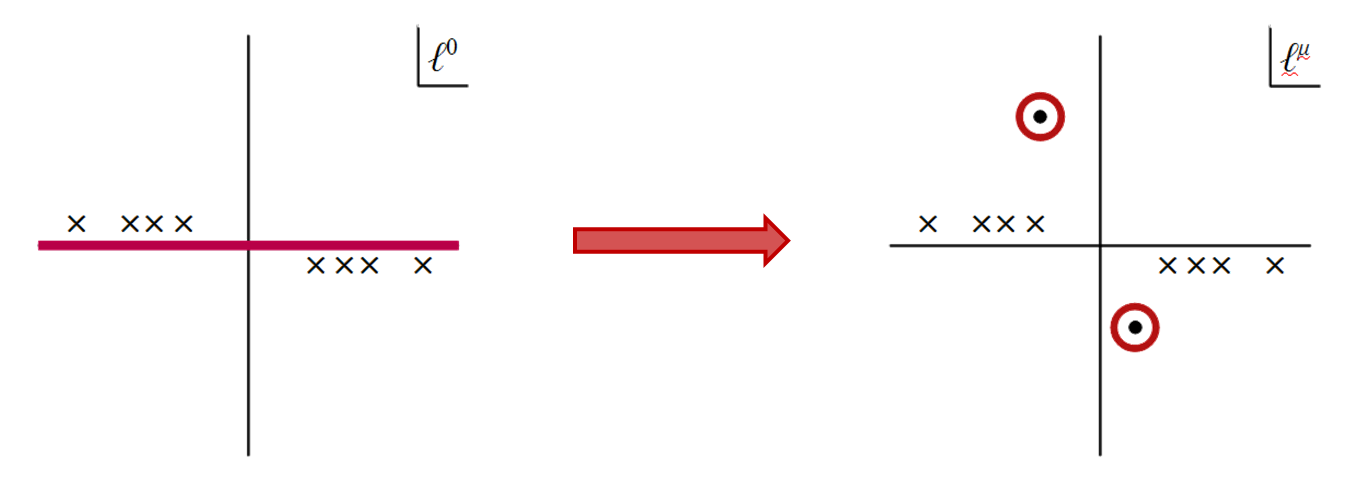}
\caption{Cutting as contour replacement
\label{ContourReplacementFigure}}
\end{figure}

That is, we must reinterpret cutting propagators as contour 
{\it replacement\/}: instead
of replacing the propagators by delta functions, we replace the
original contours of integration, along the real axes of the
now-complexified loop momenta $\ell^\mu$, by contours surrounding the
{\it global poles\/}, that is the simultaneous solutions to
eqs.~(\ref{QuadCutEquations}), in $\mathbb{C}^4$.  The contour in the case of
the one-loop box is a product of four circles, that is a four-torus
$T^4$.  The replacement is illustrated schematically in
fig.~\ref{ContourReplacementFigure}, with contours
${\cal C}_{1,2}$ encircling the two global poles.  We stress that this is {\it
  not\/} a contour deformation leaving the value of the integral
unchanged; it is a replacement, changing the value of the
integral and ultimately allows us to derive an equation for the
coefficient of the one-loop box.

This reinterpretation raises two new problems, however.  We have to
choose $T^4$ contours surrounding two global poles, around which we
could wind an arbitrary number of times (even a fractional number of
times).  How should we choose the contour?  Also, replacing the
contour by such an arbitrary winding can break integral identities.
For example, the identity,
\begin{equation}
0 = \int {d^{4-2\e}\ell\over (2\pi)^{4-2\e}}\;
{\varepsilon(\ell,k_1,k_2,k_4) \over \ell^2 (\ell-k_1)^2 (\ell-K_{12})^2 (\ell+k_4)^2}\,,
\label{EpsilonEquation}
\end{equation}
is spoiled if we choose the contour to encircle just one of the global poles.

Remarkably, these two problems cancel each other out.  If we take a
general contour, ${\cal C} = a_1 {\cal C}_1 + a_2 {\cal C}_2$, we find
that the integral in eq.~(\ref{EpsilonEquation}) takes the value,
\begin{equation}
(a_1-a_2) f(k_1,k_2,k_4)\,,
\end{equation}
so that it will still vanish if $a_1=a_2$.  This fixes the contour up
to an overall irrelevant constant which will cancel out of
eq.~(\ref{BasicEquation}).  Applying the cutting via contour
replacement to our basic equation~(\ref{BasicEquation}), we can derive
a formula for the coefficient of the corresponding one-loop box.  The
resulting equation is the same as the one obtained by Britto, Cachazo,
and Feng (BCF)~\cite{BCFUnitarity}.

Let us now apply these ideas to the double-box integral
(fig.~\ref{DoubleBoxFigure}).  In the same way that this derivation
can be seen as a generalization to two loops of the formalisms of BCF
and Forde~\cite{Forde}, recent work on two-loop integrands by
Mastrolia, Mirabella, Ossola, and Peraro~\cite{OssolaMastrolia} and by
Badger, Frellesvig, and Zhang~\cite{BadgerFZ,Zhang} can be seen as the
two-loop generalization of the Ossola--Papadopoulos--Pittau
construction~\cite{OPP}.  The maximal cut in the double box involves
cutting seven propagators.  Each solution has one continuous degree of
freedom $z$.  The number of distinct solutions depends on the number
and configuration of external masses.  When all four external legs are
massless; when one external leg is massive; when two
diagonally-opposite legs (for example, legs~1 and~3) are massive; or
when two long-edge legs (for example, legs~1 and~4) are massive, there
are six solutions.  We will call these configurations `class (c)'.
When two short-edge legs (for example, legs~1 and~2) are massive, or
when three legs are massive, there are four solutions.  We will label
these configurations `class (b)'.  (This classification is explained
in ref.~\cite{CaronHuotLarsen}.)
Performing the corresponding
contour integrals gives rise to a Jacobian factor (the analog for
the maximal cut in the double box of the
${\rm Poly}_2'(z_0)$ factor in eq.~(\ref{ContourIntegral})).  Here,
the resulting Jacobian is a function of $z$; it has poles in $z$, so
that we can choose a contour for the $z$ integration as well, thereby
obtaining a global pole.  After identifying different parametrizations
of the same global pole arising from different solutions to the
heptacut equations, and taking into account poles that arise in the
loop momenta $\ell_{1,2}$ as well as in the Jacobian, we see that
there are always eight global poles, independent of the number and
configuration of external masses~\cite{CaronHuotLarsen}.

Unlike the number of global poles,
the number of independent basis integrals does depend on the number
and configuration of external masses.  In class~(c), there are two
basis integrals; in class~(b), there are three.  We would like to
construct independent ``projectors'', which give formul\ae{} for the
coefficients of each of these basis integrals.  Each projector will be
a linear combination of contours around the global poles.  How should
we choose them?

We again impose the requirement, analogous to
eq.~(\ref{EpsilonEquation}), that all vanishing loop integrals
continue to vanish on the chosen contours.  As at one loop, there are
vanishing integrals where a Levi-Civita tensor is inserted into the
numerator of the scalar integral.  There are five possible integrals,
which give rise to four independent constraints on the contours.  In
addition, there are integration-by-parts (IBP)
identities~\cite{IBP,Laporta} which give 20 linear relations 
in class (c) between
the 22 integrals with different powers of the two irreducible
numerators $\ell_1\cdot k_4$ and $\ell_2\cdot k_1$.  Not all of the
resulting constraints on the contours are independent; in class~(c),
we find two independent constraints, while in class~(b) we have 
only 19 IBP equations, which reduce to a
single constraint.  We work here to leading order in $\e$, leaving
higher-order terms in the coefficients to future work.  The
constraints leave us with two independent contours for the two master
integrals in class~(c), and three independent contours for the three
master integrals in class~(b).  We can obtain a projector for any
given integral by imposing the further constraint that the other
integrals vanish on the contour, and that it reproduces the integral
itself with unit coefficient.

The formul\ae{} for the coefficients of double boxes take the
following form,
\begin{equation}
c = i\sum_{j=1}^8 a_j \oint_{T^8(\Global_j)} d^8v_{a,b}\;\jig{j}
\sum_{{\rm particles\atop\rm helicities}}
 \prod_{p=1}^6 A^\treelevel_p(v_{a,b})\,,
\label{BasicCoefficientEquation}
\end{equation}
where $A^\treelevel$ are tree-level amplitudes in the gauge theory,
and the $a_j$ are weights (or winding numbers) for the different
global poles.  For example, for the one-mass double box (with
$m_1^2\neq 0$) the weights $a_j$ for the two basis integrals, $I[1]$
and $I[\ell_1\cdot k_4]$, are,
\begin{equation}
(a_j) = \frac{1}4\left(1,1,1,1,0,0,1,1\right)\,,\end{equation}
and
\begin{equation}
(a_j) = \frac{m_1^2-s_{12}}{2 s_{12} s_{14}}\left(1,1,1,1,-2,-2,3,3\right)\,,
\end{equation}
respectively.
The solutions and global poles are given in terms of a parametrization of the loop momenta given in ref.~\cite{MassiveTwoLoop},
where the reader may also find complete formul\ae{} for the solutions, global poles, and projectors for classes~(b) and~(c).

This work is supported by
the European Research Council under Advanced Investigator Grant
ERC--AdG--228301.

\end{document}